\documentclass[a4,10pt]{article}

\newcommand{\trace}[1]{\ensuremath{\mathrm{Tr}\left[{#1}\right]}}
\newcommand{\partrace}[2]{\ensuremath{\mathrm{Tr}_{{#2}}\left[{#1}\right]}}

\newcommand{\func}[1]{\ensuremath{\left[ {#1} \right]}}
\begin{document}
\title{Landauer's erasure principle in non-equilibrium systems.}
\author{O.~J.~E.~Maroney \\
Faculty of Philosophy, University of Oxford\footnote{Mailing address: Wolfson College, Linton Road, Oxford, OX2 6UD, UK}\\
owen.maroney@philosophy.ox.ac.uk}
\date{\today}
\maketitle
\begin{abstract}
In two recent papers, Maroney and Turgut separately and independently show generalisations of Landauer's erasure principle to indeterministic logical operations, as well as to logical states with variable energies and entropies.  Here we show that, although Turgut's generalisation seems more powerful, in that it implies but is not implied by Maroney's and that it does not rely upon initial probability distributions over logical states, it does not hold for non-equilibrium states, while Maroney's generalisation holds even in non-equilibrium.  While a generalisation of Turgut's inequality to non-equilibrium seems possible, it lacks the properties that makes the equilibrium inequality appealing.  The non-equilibrium generalisation also no longer implies Maroney's inequality, which may still be derived independently.  Furthermore, we show that Turgut's inequality can only give a necessary, but not sufficient, criteria for thermodynamic reversibility.  Maroney's inequality gives the necessary and sufficient conditions.
\end{abstract}
\section{Introduction}
In two recent papers,\cite{Maroney2007b,Turgut2006}, Landauer's erasure principle\cite{Lan61,LR03} is extended and generalised to include indeterministic logical operations, and variable internal energy and entropy of the physical representations of logical states.  In \cite{Maroney2007b} the generalised principle is expressed in the form
\begin{eqnarray}\label{eq:maroney}
\sum_{\alpha,\beta} P(\alpha,\beta) W_{\alpha,\beta} & \geq & \sum_\beta P(\beta)\left(F_\beta+k T \ln P(\beta)\right) \nonumber \\ & & -\sum_\alpha P(\alpha)\left(F_\alpha+k T \ln P(\alpha)\right)
\end{eqnarray}
while in \cite{Turgut2006} it is expressed as
\begin{equation}\label{eq:turgut}
\sum_{\alpha} P(\beta|\alpha) \exp^{-\left(W_{\alpha,\beta} +F_\alpha -F_\beta\right)/kT} \leq 1
\end{equation}
Here $F_\alpha=E_\alpha-TS_\alpha$, where $E_\alpha$ is the mean internal energy of the logical state $\alpha$, $S_\alpha$ its entropy, $T$ is the temperature of the surroundings and $W_{\alpha,\beta}$ is the expectation value for the work performed during an $\alpha$ to $\beta$ transition.

The principal differences between the two proofs are:
\begin{enumerate}
\item Maroney's proof is derived in the context of canonical statistical mechanics, while Turgut uses microcanonical statistical mechanics;
\item Turgut's proof requires logical states to be in thermal equilibrium, while Maroney's proof also applies to states out of equilibrium;
\item Maroney's proof requires the assumption of a probability distribution over the initial logical states, while Turgut's inequality is derived without using such a probability distribution;
\item Turgut shows Eq. \ref{eq:turgut} implies, but is not implied by, Eq. \ref{eq:maroney}, so represents a stronger constraint.
\end{enumerate}
Turgut's inequality has the great advantage that it is expressed solely in terms of the properties of the physical states $\{ E_\alpha, S_\alpha, E_\beta, S_\beta\}$ and the logical operation $\{P(\beta|\alpha)\}$.  Maroney's inequality also involves the contingent probability $\{P(\alpha)\}$.  However, the microcanonical approach of Turgut requires system and environment to be in thermal equilibrium while Maroney's canonical approach allows the system to be out of equilibrium.  Recent discussions of the validity of Landauer's erasure principle (see \cite{Shenker2000,Mar05b,Nor05,SLGP05,Maroney2009,Norton2010}) have raised the question of whether probability distributions over the input logical states are an admissible tool for analysing the thermodynamic cost of computation. It is therefore important to establish if Turgut's inequality remains valid in the non-equilibrium regime.

Using canonical statistical mechanics, Maroney's approach will be shown to yield a similar inequality to Eq. \ref{eq:turgut}.  This reproduces Eq. \ref{eq:turgut} in thermal equilibrium, but will show how Turgut's inequality, Eq. \ref{eq:turgut}, can be violated for non-equilibrium states.

It will also be demonstrated that, even in thermal equilibrium, reaching the equality in Eq. \ref{eq:turgut} does not imply reaching the equality in Eq. \ref{eq:maroney}.  Reaching Maroney's inequality is a necessary and sufficient condition for thermodynamic reversibility, while, even in thermal equilibrium, reaching Turgut's inequality is necessary but not sufficient.
\section{Preliminaries}
The following statistical mechanical properties will be utilised:
\begin{enumerate}
 \item If $\rho$ is canonical equilibrium state, such that $\rho=e^{-H/kT}/\trace{e^{-H/kT}}$ then any orthogonal subensemble\footnote{$\rho=\sum_i P_i \rho_i$ with $\rho_i \rho_j=0$ for $i \neq j$.} $\rho^\prime$, of $\rho$, occurring with probability $P(\rho^\prime|\rho)$, is also a canonical state with free energy $F\func{\rho^\prime}=F\func{\rho}+kT \ln \func{P(\rho^\prime|\rho)}$;
 \item Given any two states $\rho$ and $\rho^\prime$, and a single heat bath (represented by a canonically distributed system $\rho_{HB}(T)=e^{-H_{HB}/kT}/\trace{e^{-H_{HB}/kT}}$, initially uncorrelated with any other system) the work required, $W$, to deterministically evolve $\rho$ into $\rho^\prime$ is bounded by $W \geq (E\func{\rho^\prime}-TS\func{\rho^\prime})-(E\func{\rho}-TS\func{\rho})$, where $E\func{\rho}=\trace{H \rho}$ and $S\func{\rho}=-k \trace{\rho \ln \rho}$.
 \item The general inequality
  \begin{equation}\label{eq:legsum}
   \sum_i P(i) \left( A_i - \ln P(i)\right) \leq \ln \func{\sum_i \exp^{A_i} }
  \end{equation}
  with equality occurring if, and only if, $P(i)=\exp^{A_i}/\sum_i \exp^{A_i}$.
\end{enumerate}
\section{Derivation}
A general logical operation is a map from a set of input logical states, $\{\alpha\}$, to a set of output logical states, $\{\beta\}$, with the conditional probability $P(\beta|\alpha)$ that the initial logical state $\alpha$ ends up in the final logical state $\beta$.  Logical states $\alpha$ will be physically represented by the system density matrix $\rho_\alpha$, and $\beta$ by $\rho_\beta$.  The environment is assumed to be initially in an uncorrelated canonical state $\rho_{E}=e^{-H_{E}/kT}/\trace{e^{-H_{E}/kT}}$.  The system and environment has a joint Hamiltonian $H(t)=H_S+H_E+V_{SE}$.  Physical operations are effected through manipulation of $H_S$ and heat is transferred to the environment through the interaction potential $V_{SE}$.  It is assumed that the interaction energy is negligible $\trace{V_{SE}\rho_{SE}} \ll \trace{H(t)\rho_{SE}}$.

The unitary evolution $U=\exp^{\imath \int_0^t H(t) t/\hbar}$ will be a valid physical implementation of the logical operation, if it fulfils the criteria:
\begin{eqnarray}
\rho^{\alpha}_I &=& \rho_\alpha \otimes \rho_E =\sum_\beta P(\beta|\alpha)\rho^{\alpha}_{\beta,I} \otimes \rho_E \\
\rho^{\alpha}_F &=& U \rho^{\alpha}_I U^\dag =\sum_\beta P(\beta|\alpha)U \rho^{\alpha}_{\beta,I} \otimes \rho_E U^\dag \\
\rho^{\alpha,\beta}_F &=& U \rho^{\alpha}_{\beta,I} \otimes \rho_E U^\dag \\
\rho^{\alpha}_{\beta,F} &=& \partrace{\rho^{\alpha,\beta}_F}{E} \\
\rho^{\alpha,\beta}_E &=& \partrace{\rho^{\alpha,\beta}_F}{S}
\end{eqnarray}
where the $\{\rho^{\alpha}_{\beta,I}\}$ are mutually orthogonal subensembles of the $\rho_\alpha$ density matrix, occurring with probabilities $P(\beta|\alpha)$, and where there exists a set of weights $w(\alpha|\beta) \geq 0$ such that $\sum_\alpha w(\alpha|\beta) =1$ and $\rho_\beta =\sum_\alpha w(\alpha|\beta) \rho^{\alpha}_{\beta,F}$, with the $\{\rho^{\alpha}_{\beta,F}\}$ also mutually orthogonal\footnote{Idealised physical instantiations of processes which meet this criteria are demonstrated in \cite{Maroney2007b}.  More general processes, in which the system subensembles are not necessarily mutually orthogonal, are possible utilising mutually orthogonal subensembles of the joint system and environment.  This complicates the calculation without materially affecting the conclusion.}.

Standard calculations yield heat generation and work requirements for an $\alpha$ to $\beta$ transition:
\begin{eqnarray}\label{eq:heat}
Q_{\alpha,\beta} &=&\trace{H_E \rho^{\alpha,\beta}_E -H_E\rho_E} \geq T \left(S\func{\rho^{\alpha}_{\beta,I}}-S\func{\rho^{\alpha}_{\beta,F}}\right) \\
W_{\alpha,\beta} &=& E\func{\rho^{\alpha}_{\beta,F}} -E\func{\rho^{\alpha}_{\beta,I}}+Q_{\alpha,\beta}
\end{eqnarray}
Defining the quantities:
\begin{eqnarray}
k T \ln f(\beta|\alpha) &=& \left(E\func{\rho_\alpha}-TS\func{\rho_\alpha}\right) \nonumber \\ & & -\left(E\func{\rho^{\alpha}_{\beta,I}}-TS\func{\rho^{\alpha}_{\beta,I}}\right) \nonumber \\ &=& F\func{\rho_\alpha} -F\func{\rho^{\alpha}_{\beta,I}} \\
k T \ln q(\alpha|\beta) &=& \left(E\func{\rho_\beta}-TS\func{\rho_\beta}\right) \nonumber \\ & & -\left(E\func{\rho^{\alpha}_{\beta,F}}-TS\func{\rho^{\alpha}_{\beta,F}}\right) \nonumber \\ &=& F\func{\rho_\beta} -F\func{\rho^{\alpha}_{\beta,F}}
\end{eqnarray}
where for brevity we write $F\func{\rho}=E\func{\rho}-TS\func{\rho}$ (this is identical to the free energy for a system in a canonical equilibrium state) and we define $\mathcal{W_{\alpha,\beta}}= \left( W_{\alpha,\beta} -F\func{\rho_\alpha} +F\func{\rho_\beta} \right) /kT$ to get
\begin{equation}\label{eq:noneqsingle}
\ln f(\beta|\alpha) - \mathcal{W_{\alpha,\beta}} \leq \ln q(\alpha|\beta)
\end{equation}
giving
\begin{equation}\label{eq:noneq}
\sum_\alpha f(\beta|\alpha) \exp^\mathcal{-W_{\alpha,\beta}} \leq \sum_\alpha q(\alpha|\beta)
\end{equation}
This represents the generalisation of Turgut's inequality (Eq. \ref{eq:turgut}) to non-equilibrium states using the canonical statistical mechanics of \cite{Maroney2007b}.
\section{Equilibrium}
If the input state $\rho_\alpha$ is a canonical equilibrium state $\rho_\alpha=e^{-H/kT}/\trace{e^{-H/kT}}$, then $P(\beta|\alpha)=f(\beta|\alpha)$.

It can also be readily seen that
\begin{equation}
\sum_\alpha q(\alpha|\beta) = \sum_\alpha \exp^{-F\func{\rho^{\alpha}_{\beta,F}}/kT}/\exp^{-F\func{\rho_\beta}/kT}
\end{equation}
Using the inequality Eq. \ref{eq:legsum}, we have
\begin{eqnarray}
F\func{\rho_\beta} & = & \sum_\alpha w(\alpha|\beta) \left( F\func{\rho^{\alpha}_{\beta,F}} +kT \ln w(\alpha|\beta)  \right) \\
 & \geq & -k T \ln \func{\sum_\alpha \exp^{-F\func{\rho^{\alpha}_{\beta,F}}/kT} }
\end{eqnarray}
so $\sum_\alpha q(\alpha|\beta) \geq 1$ with equality occurring if, and only if, $w(\alpha|\beta)=\exp^{-F\func{\rho^{\alpha}_{\beta,F}}/kT} /  \sum_\alpha \exp^{-F\func{\rho^{\alpha}_{\beta,F}}/kT}$.  If this is the case the output state $\rho_\beta$ is distributed in canonical equilibrium with respect to its $\rho^{\alpha}_{\beta,F}$ subensembles, and $q(\alpha|\beta)=w(\alpha|\beta)$.

If the input states $\rho_\alpha$ and output states $\rho_\beta$ are canonical equilibrium states Eq. \ref{eq:noneq} becomes
\begin{equation}\label{eq:canoneq}
\sum_\alpha P(\beta|\alpha) \exp^\mathcal{-W_{\alpha,\beta}} \leq 1
\end{equation}
reproducing Turgut's inequality Eq. \ref{eq:turgut} in the canonical approach to statistical mechanics.
\section{Non-equilibrium}
If either the input states or the output states (or both) are non-equilibrium states, it is possible to violate Turgut's inequality.  The inequality in Eq. \ref{eq:noneq} comes from the inequality in Eq. \ref{eq:heat}.  In the limiting case of quasistatic reversible processes, the equality can be reached\footnote{Again, see \cite{Maroney2007b} for idealised examples.}.  In this section, we will assume we are in these limiting cases, so that
\begin{equation}\label{eq:noneqlimit}
\sum_\alpha f(\beta|\alpha) \exp^\mathcal{-W_{\alpha,\beta}} = \sum_\alpha q(\alpha|\beta)
\end{equation}
\subsection{Input states}
Suppose that for a particular $\beta$, $\rho_\beta$ is a canonical equilibrium state, so that:
\begin{equation}\label{eq:canonouteq}
\sum_\alpha f(\beta|\alpha) \exp^\mathcal{-W_{\alpha,\beta}} = 1
\end{equation}
If any of the states $\rho_\alpha$ are non-equilibrium, then it is always possible to find a subensemble $\rho^\alpha_\beta$ that occurs with $P(\beta|\alpha) > f(\beta|\alpha)$.
\begin{equation}
\sum_\alpha P(\beta|\alpha) \exp^\mathcal{-W_{\alpha,\beta}}>  1
\end{equation}
follows immediately.

Note that it is not possible to do this simultaneously for every possible $\beta$ output state, as $\sum_\beta f(\beta|\alpha) \geq 1$.  Turgut's inequality can therefore only be violated this way for transitions to a subset of the output states.
\subsection{Output states}
Suppose that all the input states are canonical equilibrium states, so that:
\begin{equation}\label{eq:canoneqin}
\sum_\alpha P(\beta|\alpha) \exp^\mathcal{-W_{\alpha,\beta}} =\sum_\alpha q(\alpha|\beta)
\end{equation}.
If the output state $\rho_\beta$ is not a canonical equilibrium state, then $\sum_\alpha q(\alpha|\beta)>1$ and once again:
\begin{equation}
\sum_\alpha P(\beta|\alpha) \exp^\mathcal{-W_{\alpha,\beta}} >1
\end{equation}.
In this case, Turgut's inequality may be separately violated for every single $\beta$ output state.
\section{Dependencies}
We now consider the relationship between the non-equilibrium generalisation of Turgut's inequality, and Maroney's inequality.

First let us note that, in order for the process to properly produce $\rho_\beta$, it is necessary that $\rho^{\alpha}_{\beta,F}$ occur with conditional probability $P(\alpha|\beta)=w(\alpha|\beta)$.  As
$P(\alpha|\beta)=\frac{P(\beta|\alpha)P(\alpha)}{\sum_\alpha P(\beta|\alpha)P(\alpha)}$
this will generally depend upon the input probabilities\footnote{For logically reversible operations $P(\alpha|\beta) \in \{0,1\}$ and so this is easily satisfied instead by $\rho^{\alpha}_{\beta,F}=\rho_\beta$}.  so requires $U$ to be tailored to specific input probability distributions.

If $P(\alpha|\beta) \neq w(\alpha|\beta)$, then further steps may be necessary to reach $\rho_\beta$.  One would be to allow the density matrix $\rho^\prime_\beta=\sum_\alpha P(\alpha|\beta)\rho^{\alpha}_{\beta,F}$ equilibrate to a density matrix $\rho^{\prime\prime}_\beta$ which can then be evolved into $\rho_\beta$.  The only candidate density matrix to which all possible $\rho^\prime_\beta$ matrices will equilibrate is a canonical equilibrium density matrix $\rho^{\prime\prime}_\beta=e^{-H_S/kT}/\trace{-H_S/kT}$.  This process entails an additional work requirement
\begin{equation}\label{eq:workexcess}
\Delta W_\beta=F\func{\rho^\prime_\beta}-F\func{\rho^{\prime\prime}_\beta}\geq 0
 \end{equation}
 which is minimised if and only if $\rho^\prime_\beta=\rho^{\prime\prime}_\beta$. Again the minimal work requirement only occurs for a specific input probability distribution\footnote{Note that this additional procedure is sufficient to ensure that $\sum_\alpha f(\beta|\alpha) \exp^\mathcal{-W_{\alpha,\beta}} \leq 1$, but not to ensure that $P(\beta|\alpha)\leq f(\beta|\alpha)$.}.

Averaging over the input probability distributions, Eq. \ref{eq:noneqsingle} gives
\begin{equation}
\sum_{\alpha,\beta}P(\alpha,\beta)\left( kT \ln f(\beta|\alpha) - \mathcal{W_{\alpha,\beta}} - kT \ln q(\alpha|\beta) \right) \geq 0
\end{equation}
Using the optimal process $w(\alpha|\beta)=P(\alpha|\beta)$ and the identities
\begin{eqnarray}
\sum_\alpha w(\alpha|\beta) \ln \frac{w(\alpha|\beta)}{q(\alpha|\beta)}&=&0 \label{eq:qident}\\
\sum_\beta P(\beta|\alpha) \ln \frac{P(\beta|\alpha)}{f(\beta|\alpha)}&=&0 \label{eq:fident}\\
\end{eqnarray}
Maroney's inequality Eq. \ref{eq:maroney} follows.

To understand the relationship with Eq. \ref{eq:noneq}, consider Eq. \ref{eq:legsum} with $P(i)=P(\alpha|\beta)$ and $x_i=-\mathcal{W_{\alpha,\beta}} + \ln f(\beta|\alpha)$
\begin{eqnarray}
\sum_\alpha & P(\alpha|\beta) &\left( \mathcal{W_{\alpha,\beta}} - \ln f(\beta|\alpha)+\ln P(\alpha|\beta) \right) \nonumber \\ && \geq -\ln \sum_\alpha f(\beta|\alpha) \exp^\mathcal{-W_{\alpha,\beta}}
\end{eqnarray}

In equilibrium $f(\beta|\alpha)=P(\beta|\alpha)$ which with Turgut's inequality gives
\begin{equation}
\sum_\alpha P(\alpha|\beta)\left( \mathcal{W_{\alpha,\beta}} - \ln f(\beta|\alpha)+\ln P(\alpha|\beta) \right) \geq 0
\end{equation}
Multiplying by $P(\beta)$ and summing gives Maroney's inequality.

However, out of equilibrium we have
\begin{eqnarray}
 -\ln \sum_\alpha f(\beta|\alpha) \exp^\mathcal{-W_{\alpha,\beta}} \geq -\ln \sum_\alpha q(\alpha|\beta) \leq 0
\end{eqnarray}
so Eq. \ref{eq:noneq} does not directly imply Eq. \ref{eq:maroney}.
\section{Thermodynamic Reversibility}
There are four inequalities:
\begin{enumerate}
\item Eq. \ref{eq:heat}, with equality reached in the limit of quasistatic processes, in which case \begin{equation}\label{eq:eqqsl}
    q(\alpha|\beta)=f(\beta|\alpha) e^\mathcal{-W_{\alpha,\beta}}
    \end{equation}
\item Eq. \ref{eq:legsum}, with equality reached for the probabilities
\begin{equation}\label{eq:eqprob}
P(\alpha|\beta)=\frac{f(\beta|\alpha) e^\mathcal{-W_{\alpha,\beta}}}{\sum_\alpha f(\beta|\alpha) e^\mathcal{-W_{\alpha,\beta}}}
\end{equation}
\item The non-equilibrium inequality for the output states $\sum_\alpha q(\alpha|\beta) \geq 1$ with equality reached for equilibrium output states:
    \begin{equation}\label{eq:eqeq}
    w(\alpha|\beta)=q(\alpha|\beta)
    \end{equation}
\item Eq. \ref{eq:workexcess}, which may be minimised by optimising the unitary evolution $U$, for the input probabilities $P(\alpha)$ such that:
\begin{equation}\label{eq:eqopt}
P(\alpha|\beta)=w(\alpha|\beta)
\end{equation}
\end{enumerate}
To reach equality in Eq. \ref{eq:maroney}, it is necessary and sufficient that both Eq. \ref{eq:eqqsl} and Eq. \ref{eq:eqopt} hold, while to reach the equality in Eq. \ref{eq:noneq} it is necessary and sufficient that only Eq. \ref{eq:eqqsl} holds.  Reaching the equality in Eq. \ref{eq:noneq} is necessary but not sufficient for the equality in Eq. \ref{eq:maroney}.  As we have argued in detail\cite{Maroney2007b}, meeting the equality in Eq. \ref{eq:maroney} represents thermodynamic reversibility.  We also note that equilibrium, Eq. \ref{eq:eqeq}, is neither necessary nor sufficient.

Finally, we will note that, although $-\ln \sum_\alpha q(\alpha|\beta) \leq 0$ this does not lead to the possibility that $\sum_\alpha P(\alpha|\beta)\left( \mathcal{W_{\alpha,\beta}} - \ln f(\beta|\alpha)+\ln P(\alpha|\beta) \right) <0$.  Satisfying conditions Eq. \ref{eq:eqqsl} and Eq. \ref{eq:eqprob} implies
$P(\alpha|\beta)=\frac{q(\alpha|\beta)}{\sum_\alpha q(\alpha|\beta)}$, but then the identity Eq. \ref{eq:qident} with Eq. \ref{eq:eqopt} ensures that condition Eq. \ref{eq:eqeq} also holds.  (We also note in passing that Eq. \ref{eq:eqqsl}, Eq. \ref{eq:eqeq} and Eq. \ref{eq:eqopt} together imply the condition Eq. \ref{eq:eqprob} holds).
\section{Conclusion}
The usefulness of Turgut's inequality lies in the fact that it is defined entirely in terms of properties of the logical operation, and the physical states, and has no contingent factors.  Unfortunately this inequality does not hold in general out of equilibrium.  Generalising it to non-equilibrium is possible, but Eq. \ref{eq:noneq} introduces factors such as $f(\beta|\alpha)$ and $q(\alpha|\beta)$.  Although $\sum_\beta f(\beta|\alpha)$ and $\sum_\alpha q(\alpha|\beta)$ are, to some extent, measures of how far $\rho_\alpha$ and $\rho_\beta$ are `out-of-equilibrium', the specific values $f(\beta|\alpha)$ and $q(\alpha|\beta)$ are contingent upon the specific physical instantiation of the logical operation.

Maroney's inequality also contains contingent factors - the probabilities with which the input logical states occur.  Consideration of these probabilities turns out to be essential to thermodynamic reversibility, at least for logically irreversible operations.

Although in equilibrium Turgut's inequality implies Maroney's inequality, the generalised inequality does not imply Maroney's inequality out of equilibrium.  Instead, both Eq. \ref{eq:noneq} and Eq. \ref{eq:maroney} may be independently derived from the same assumptions.  With regard to reaching the equality, the necessary and sufficient conditions for reaching Eq. \ref{eq:noneqlimit} (namely, that the equality in Eq. \ref{eq:heat} is met for all transitions) are also necessary, but not sufficient for reaching the equality in Eq. \ref{eq:maroney}.  The additional conditions required are conditions regarding probability distributions over the input states, for logically irreversible operations.
\bibliographystyle{alpha}

\end{document}